\documentclass[twocolumn,superscriptaddress,showpacs]{revtex4}
\usepackage{graphicx}
\usepackage{amsmath}

\begin{document}

\title{Square lattice site percolation at increasing ranges of neighbor interactions}

\author{Krzysztof Malarz}
\homepage{http://home.agh.edu.pl/malarz/}
\affiliation{
AGH University of Science and Technology,
Faculty of Physics and Applied Computer Science\\
al. Mickiewicza 30, PL-30059 Krak\'ow, Poland
}

\affiliation{
Universit\'e Pierre et Marie Curie et CNRS,
Laboratoire des Milieux D\'esordonn\'es et H\'et\'erog\`enes\\
Case 86, 4 place Jussieu, F-75252 Paris Cedex 05, France 
}

\author{Serge Galam}
\email{galam@ccr.jussieu.fr}
\affiliation{
Universit\'e Pierre et Marie Curie et CNRS,
Laboratoire des Milieux D\'esordonn\'es et H\'et\'erog\`enes\\
Case 86, 4 place Jussieu, F-75252 Paris Cedex 05, France 
}

\date{\today}

\begin{abstract}
We report site percolation thresholds for square lattice with 
neighbor interactions at various increasing ranges. Using Monte Carlo 
techniques  we  found that  nearest neighbors (N$^2$), next nearest 
neighbors (N$^3$), next next nearest neighbors (N$^4$) and fifth 
nearest neighbors  (N$^6$) yield the same $p_c=0.592\cdots$. At odds, 
fourth nearest neighbors (N$^5$) give $p_c=0.298\cdots$. These results are 
given an explanation in terms of symmetry arguments. We then consider 
combinations of various ranges of interactions with (N$^2$+N$^3$), 
(N$^2$+N$^4$), (N$^2$+N$^3$+N$^4$) and (N$^2$+N$^5$). The calculated 
associated thresholds are respectively $p_c=0.407\cdots,\ 0.337\cdots,\ 0.288\cdots,\ 
0.234\cdots$. The existing Galam--Mauger universal formula for percolation 
thresholds  does not reproduce the data showing dimension and 
coordination number are not sufficient to build a universal law which 
extends to complex lattices.
\end{abstract}



\maketitle

\section{Introduction}

Calculating percolation thresholds has been an ongoing challenge for 
decades \cite{sykes,kesten,bunde,intro,sahimi,ds-rev}. 
While very few lattices allow an exact analytical 
calculation, large scale simulations have been very valuable to 
determine a large spectrum of them for both Bravais'
\cite{intro,suding,rosowsky}
and disordered \cite{network} lattices.
The drastic increase in computer capacities has recently 
permitted the calculation of thresholds at rather high dimensions up 
to $d=13$ for the hypercube \cite{intro,pcd13}. In parallel, not much work has been 
devoted to regular lattices with neighbor interactions which are not 
nearest neighbors (N$^2$, von Neumann's neighborhood). Some scarce 
results are available for simultaneous nearest and next nearest 
neighbors (N$^2$+N$^3$, Moore's neighborhood) \cite{intro,bulg}.

In this paper we 
report for the first time a systematic calculation of site 
percolation thresholds for the square lattice with neighbor 
interactions at successive increasing range. We consider the series 
of nearest neighbors (N$^2$), next nearest neighbors (N$^3$), next 
next nearest neighbors (N$^4$), fourth nearest neighbors (N$^5$) and 
fifth nearest neighbors (N$^6$).
It should be stressed that for each one of the considered distance of interaction, all others are not active.
For instance in the case of next nearest neighbors (N$^3$), the nearest neighbors (N$^2$) sites are not connected, only the N$^3$ are.
This principle applies to all our calculations. 
We found that the threshold is the 
same for all of them with $p_c=p_c(\text{N}^2)$ except at N$^5$.
An explanation in terms of symmetry is provided.

We then 
consider combinations of various ranges of interactions with 
(N$^2$+N$^3$), (N$^2$+N$^4$), (N$^2$+N$^5$) and (N$^2$+N$^3$+N$^4$).
In these cases we have simultaneous range of interactions but they are necessary compact.
For instance for (N$^2$+N$^4$) all nearest neighbors sites are connected as well as all next next nearest neighbor ones but next nearest neighbors are not interacting. 

Comparing our numerical estimates with 
the predictions from the Galam--Mauger (GM) universal formula for 
percolation thresholds \cite{pcdz}, we found significant discrepancies.  It 
strengthens the earlier claim that only dimension and coordination 
number could not be sufficient to build a universal law which extends 
to complex lattices \cite{suding,marck}.

\section{Calculations}

\begin{figure}
\begin{center}
(a) \includegraphics[scale=.35]{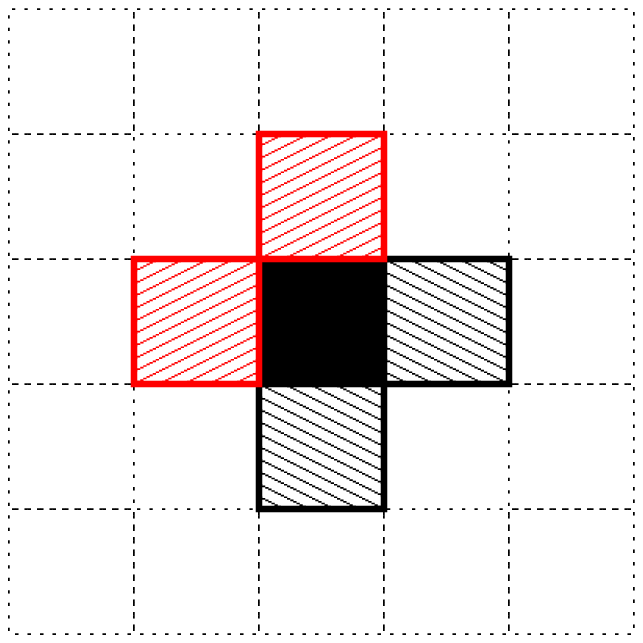}
(b) \includegraphics[scale=.35]{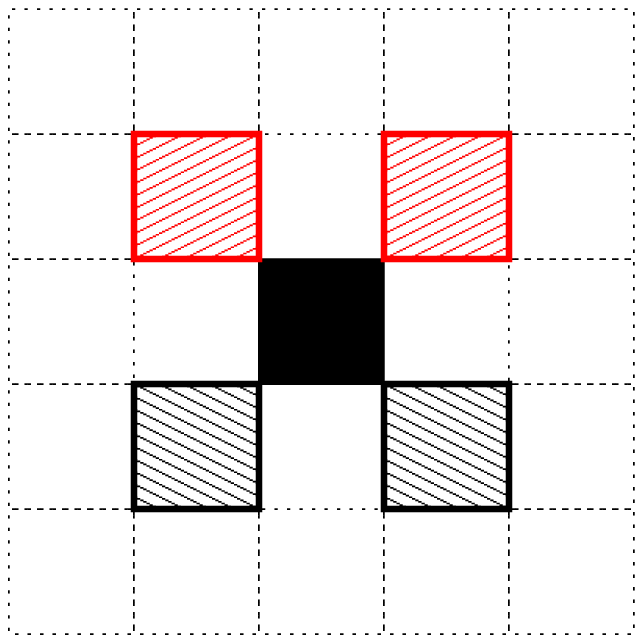}
(c) \includegraphics[scale=.35]{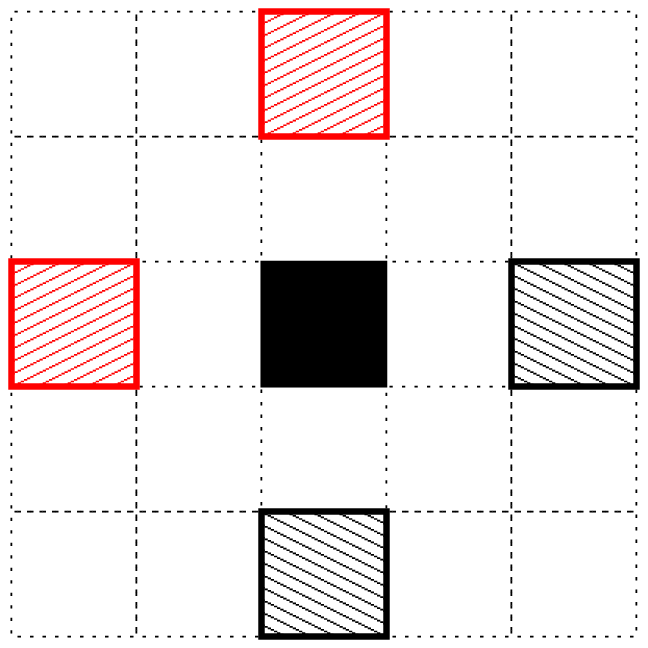} \\
(d) \includegraphics[scale=.35]{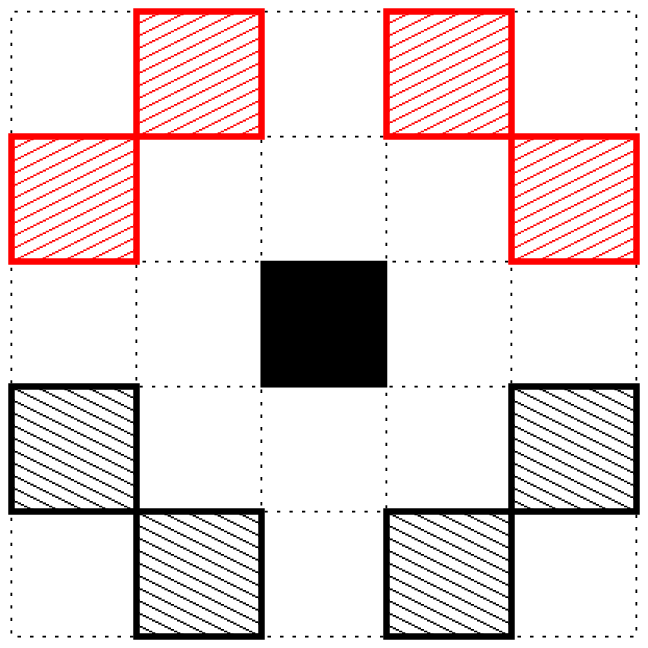}
(e) \includegraphics[scale=.35]{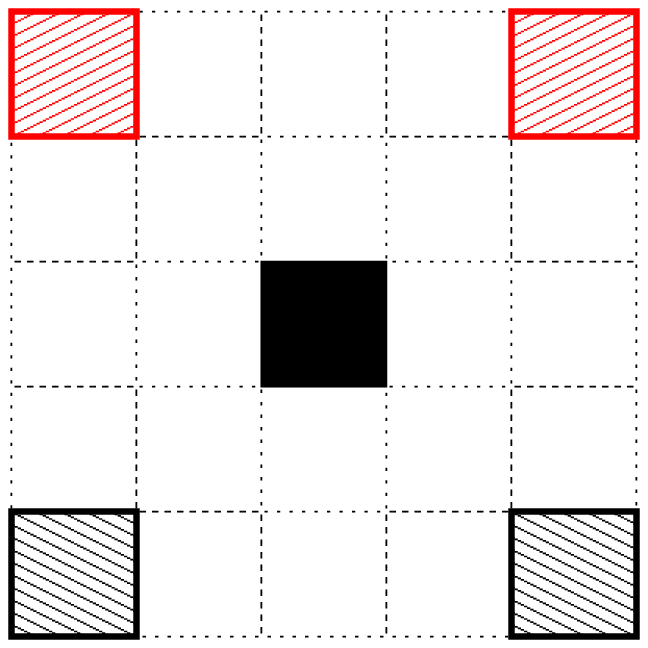}
(f) \includegraphics[scale=.35]{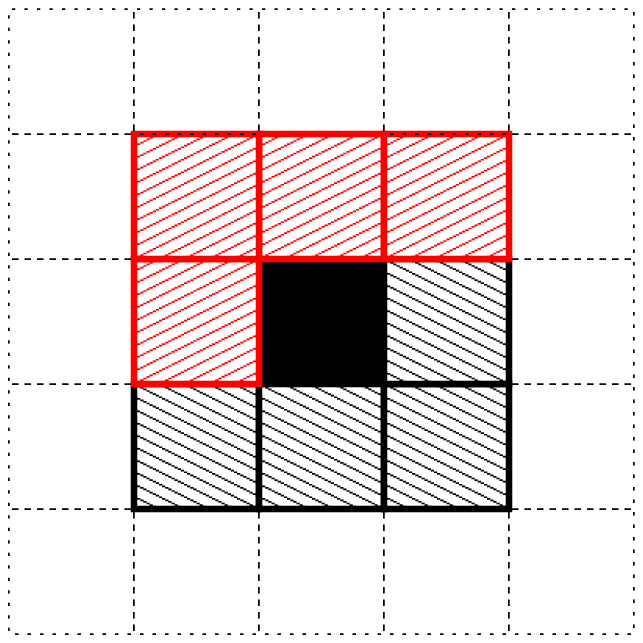} \\
(g) \includegraphics[scale=.35]{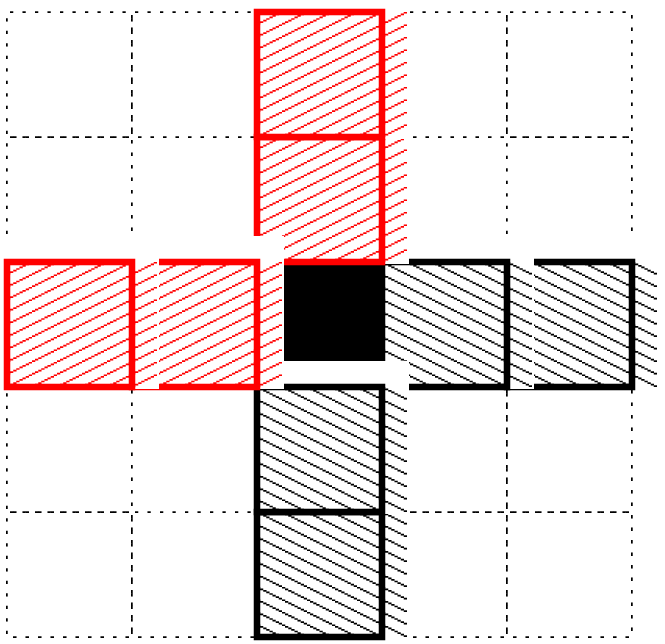}
(h) \includegraphics[scale=.35]{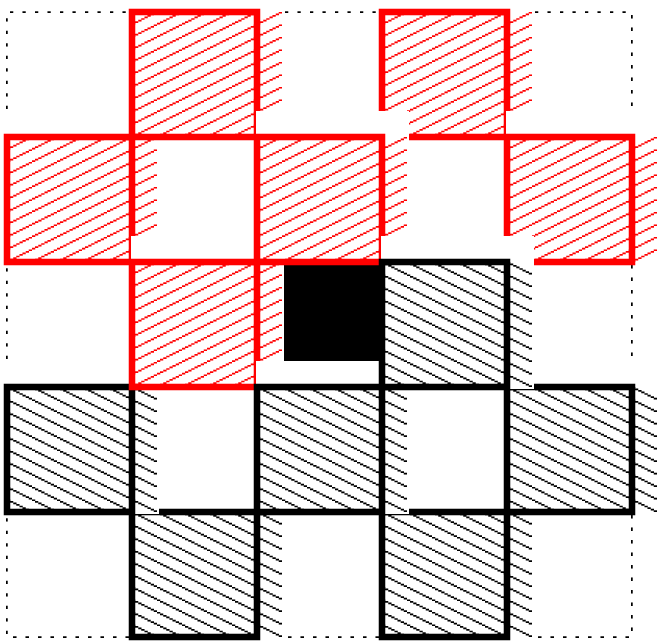}
(i) \includegraphics[scale=.35]{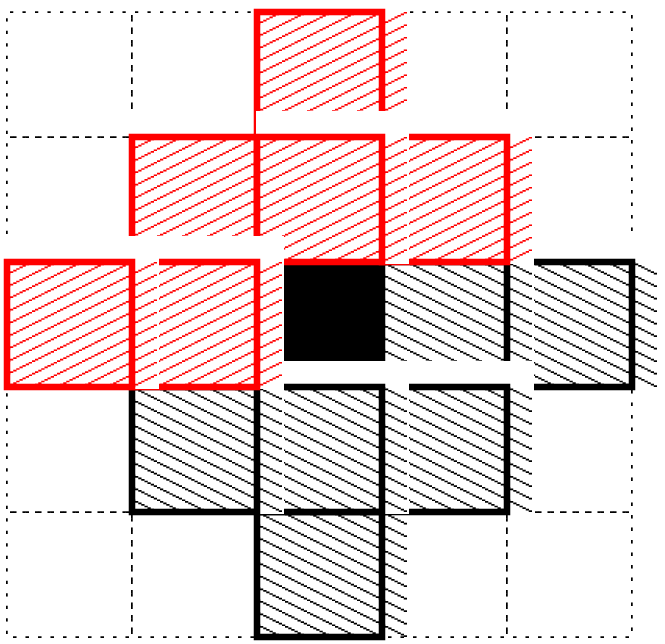}
\caption{Various site neighborhoods on the square lattice:
(a) N$^2$ --- von Neumann's neighborhood,
(b) N$^3$, (c) N$^4$, (d) N$^5$, (e) N$^6$,
and the examples of their combinations:
(f) N$^2$+N$^3$ --- Moore's neighborhood,
(g) N$^2$+N$^4$,
(h) N$^2$+N$^5$ and
(i) N$^2$+N$^3$+N$^4$.}
\label{fig-ngbr}
\end{center}
\end{figure}

There exist several computational techniques 
which allow to perform calculations of percolation thresholds 
\cite{newman,ziff,leath,hka}. Here we are using  the Hoshen--Kopelman 
algorithm (HKA) \cite{hka}. Once the lattice is given with the 
occupied sites, it allows to recognize which sites belong to which 
clusters.
With HKA one can assign to each occupied site a label and sites in 
the same cluster
have the same labels. Different labels are assigned to different clusters.
The HKA is particularly efficient when we check if the site at distance $\ell$
from the first line --- often fully occupied --- is still connected 
to that line through the sites at the distances smaller than $\ell$.
The algorithm requires storing only single line of sites and goes through
the lattice only once.
In such a case HKA becomes extremely efficient as it saves memory and time \cite{intro}. 
However, when links between sites at distances larger
than $\ell$ from a top border are desired, the whole lattice must be stored \cite{intro, km-amv, ds-aa, km-sk-kk}.  
With the HKA on a square lattice when we assign the labels for the
investigated site (black sites in Fig. \ref{fig-ngbr}), we need to check already
labeled and occupied sites in its neighborhood (slashed sites in Fig. \ref{fig-ngbr}).
The possible links to remaining sites in the neighborhood (backslashed sites
in Fig. \ref{fig-ngbr}) may be checked later, basing on the neighborhood's point symmetry.

The percolation thresholds values $p_c$ are evaluated from the 
crossing point of
three curves showing dependences of the percolation probability $P$ on the site
occupation probability $p$ for lattices of linear sizes $L=100$, 500 and 1000.
The results are averaged over $N_{\text{run}}=10^3$ and $10^4$ for $L=1000$ and
100, respectively.
With enlarging the lattice size $L$ the curve $P(p)$ becomes stepper 
and stepper
and tends to Hevisade's function $\Theta(-p_c)$ when $L\to\infty$, as expected.

\section{Results}

We present our results in Tab. \ref{tab-pc}.
\begin{table}
\caption{The percolation threshold $p_c$ for various 
neighborhoods on square lattice and sites coordination number $z$ and 
the theoretical values $p_c^\text{GM}$.}
\begin{center}
\label{tab-pc}
\begin{ruledtabular}
\begin{tabular}{llll} 	
neighborhood    & $z$& $p_c$         & $p_c^{\text{GM}}$ \\
  \hline
N$^2$            &  4 & $0.592\cdots$ & $0.5984\cdots$ \\
N$^3$            &  4 & $0.592\cdots$ & $0.5984\cdots$ \\
N$^4$            &  4 & $0.592\cdots$ & $0.5984\cdots$ \\
N$^5$            &  8 & $0.298\cdots$ & $0.4411\cdots$ \\
N$^6$            &  4 & $0.592\cdots$ & $0.5984\cdots$ \\
  \hline
N$^2$+N$^3$      &  8 & $0.407\cdots$ & $0.4411\cdots$ \\
N$^2$+N$^4$      &  8 & $0.337\cdots$ & $0.4411\cdots$ \\
N$^2$+N$^5$      & 12 & $0.234\cdots$ & $0.3748\cdots$ \\
  \hline
N$^2$+N$^3$+N$^4$& 12 & $0.288\cdots$ & $0.3748\cdots$ \\
\end{tabular}
\end{ruledtabular}
\end{center}
\end{table}
The percolation thresholds $p_c$ for the square
lattice are computed with HKA for a series of neighborhoods. First 
only one type of neighbor interactions is considered at  a time, 
increasing repeatedly the range with N$^2$, N$^3$, N$^4$, N$^5$ and 
N$^6$. It turns out that the threshold $p_c=0.592\cdots$  is the same for 
all of them except at N$^5$ where $p_c=0.298\cdots$.  

\begin{figure}
\begin{center}
(a) \includegraphics[scale=.4]{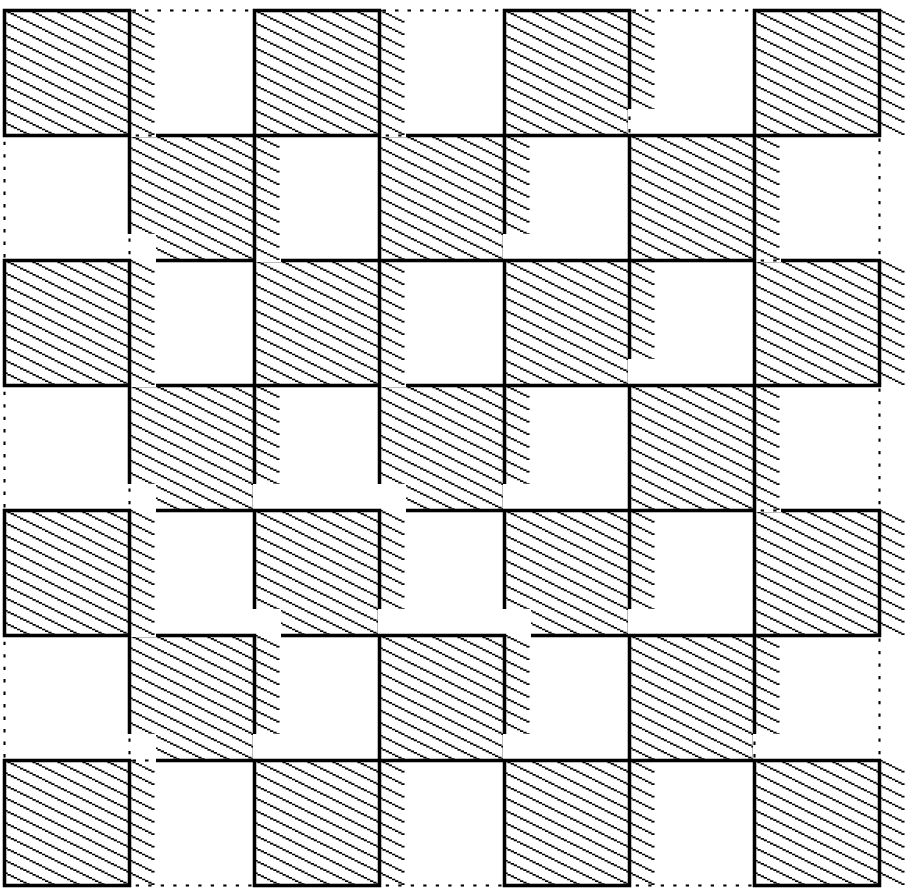}
(b) \includegraphics[scale=.4]{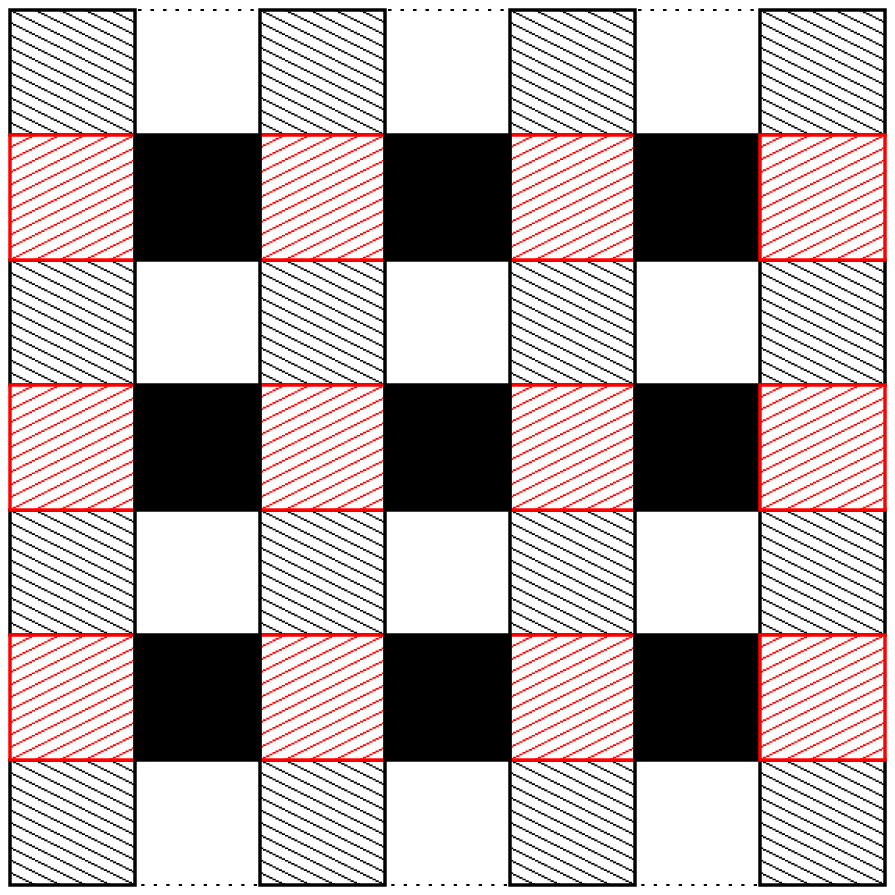}
\caption{The lattices with (a) N$^3$ and (b) N$^4$ neighborhoods may be mapped to (a) two or (b) four parallel N$^2$ situations but with (a) $\sqrt{2}$ and (b) two times larger lattice constants.}
\label{fig-map2nn}
\end{center}
\end{figure}

Indeed, all 
lattices with neighborhoods shown in Figs. \ref{fig-ngbr}(b), (c) and 
(e) may be mapped into a N$^2$ situation as in Fig. \ref{fig-ngbr}(a). 
The only difference is
a larger and larger lattice constant. To implement the mapping, we 
take a square lattice and build on it the lattice from only N$^3$ 
interactions.
Two independent interpenetrated
squares sublattices appear. Therefore the percolation of N$^3$ is 
split onto two parallel N$^2$ problems
on each one of these two square sublattices.  Accordingly
the $p_c$ on each one is the $p_c$ of N$^2$. Moreover,  as the site 
must be distributed
homogeneously on the initial lattice, we will have the same density 
of occupied sites on each one of
  the sublattice making both percolation to  occur simultaneously at 
the same $p_c$ (see Fig. \ref{fig-map2nn}(a)). 
Such a scheme can be repeated with N$^4$ (Fig. \ref{fig-map2nn}(b)) and N$^6$ 
but not with N$^5$.
As shown in Fig. \ref{fig-ngbr}(d) the N$^5$ lattice has eight 
neighbors while N$^2$, N$^3$, N$^4$ and N$^6$ have four.

These symmetry properties may become instrumental in 
underlining interesting physical properties associated to some exotic 
materials.
In particular if one is able to discriminate between the two interpenetrated lattices,
it may open a way to reach percolation at a much lower critical density, down to halt the value of $p_c$.
But such a search is out the scope of the present 
work.

We also consider several combinations of various ranges of interactions.
First an increasing compact neighborhood  with (N$^2$+N$^3$) and (N$^2$+N$^3$+N$^4$).
The calculated threshold numerical estimates are respectively $p_c=0.407\cdots$ and $0.288\cdots$ (Tab. \ref{tab-pc}).
Then more complex ones with (N$^2$+N$^4$) and (N$^ 2$+N$^5$) for which we obtained $p_c=0.337\cdots$
and $0.234\cdots$ (Tab. \ref{tab-pc}). 
The fact that $p_c$ of (N$^2$+N$^5$) is smaller than $p_c$ of (N$^2$+N$^3$+N$^4$) 
is consistent with N$^5$ $z=8$ instead of $z=4$ for all the others.

The obtained percolation threshold values $p_c(\text{N}^2)=0.592\cdots$ and
$p_c(\text{N}^2+\text{N}^3)=1-p_c(\text{N}^2)=0.407\cdots$ are consistent with the values reported
in Refs. \onlinecite{intro,bulg,pcNN}.
We have also revised the value of $p_c(\text{N}^2+\text{N}^3+\text{N}^4)$ which has been studied in an earlier paper \cite{dd} and it was put at $0.292$ to compare to our value of $0.288\cdots$.

\section{Discussion}
At this stage it is interesting to check the validity GM universal formula for percolation thresholds \cite{pcdz,wier} in the case of these complex neighbor interactions.
Comparing our numerical estimates with its predictions as shown in Tab. \ref{tab-pc} we found a good agreement for N$^2$, N$^3$, N$^4$ and N$^6$ ($\Delta= 0.006$) but not for N$^5$ ($\Delta= 0.123$).
It is also fair for (N$^2$+N$^3$) with $\Delta= 0.034$ but not for all others combinations.
The significant discrepancies occur for complex and non compact neighborhood.
It strengthens the earlier claim that only dimension and coordination number could not be sufficient to build a universal law which extends to complex lattices \cite{marck}.

Indeed above failures could be anticipated due to the fact that several lattices have both identical $z$ and $d$ tough they exhibit different thresholds as seen from Tab. \ref{tab-pc}. 
In particular N$^5$, (N$^2$+N$^3$), (N$^2$+N$^4$) have $z=8$ and $d=2$ while all
$p_c$ are different. The same occurs for (N$^2$+N$^5$) and (N$^2$+N$^3$+N$^4$)
with $z=12$ and $d=2$.

Similar situation occurs for $T_C$ in the Ising model where even with 
the same number of interacting spins in the neighborhood and the same
dimensionality we have different $T_C$ \cite{solomon}. On the other hand 
Bragg--Williams approximation \cite{bragg} predicts $T_C$ to be unique 
function of coordination number $z$, i.e. $k_BT_C=zJ$ \cite{huang}. 
The GM universal formula which also extends to $T_C$ includes 
a dependence on both $d$ and $z$ \cite{Tc_alld}.

To conclude we have report for the first time numerical estimates for site percolation 
thresholds for the square lattice with N$^3$, N$^4$, N$^5$, N$^6$,
(N$^2$+N$^4$) and (N$^2$+N$^5$) interactions.
Our new estimates may prove useful in the search for a robust universal formula for percolation thresholds which would apply to complex lattices.
In particular on how to extend the GM law by including some additional topological ingredient besides coordination $z$ and dimension $d$.

These results may prove useful to some of the large spectrum of physical and 
interdisciplinary topics where the percolation theory may be applied 
like forest fires spreading \cite{km-sk-kk,fires}, 
immunology \cite{immun}, 
liquid migration in porous media \cite{porous}, 
econophysics \cite{econo},
and sociophysics \cite{socio}.

\begin{acknowledgments}
We would like to thank UnProf. Dr. habil Dipl.-Phys. D.~Stauffer for introducing us each other during 
the 18$^{\text{th}}$ Max Born Symposium in L\c{a}dek Zdr\'oj, 2003. 
KM's stay at the Universit\'e Pierre et Marie Curie et CNRS, LMDH was financed by
the European Science Foundation with grant No. COST-\-STSM-\-P10-\-00262.
The numerical calculations were carried out in ACK\---CY\-F\-RO\-NET\---AGH.
The machine time on SGI 2800 is financed by the Polish Ministry of Science 
and Information Technology under grant No. MNiI/\-SGI2800/\-AGH/\-049/\-2003. 
\end{acknowledgments}



\begin{thebibliography}{88}

\bibitem{sykes} 
M.~F.~Sykes, M.~Glen,
J. Phys. A \textbf{9}, 87 (1976);
M.~F.~Sykes, D.~S.~Gaunt, M.~Glen,
\textit{ibid.} \textbf{9}, 97 (1976);
M.~F.~Sykes, D.~S.~Gaunt, M.~Glen,
\textit{ibid.} \textbf{9}, 715 (1976);
M.~F.~Sykes, D.~S.~Gaunt, M.~Glen,
\textit{ibid.} \textbf{9}, 725 (1976);
D.~S.~Gaunt, M.~F.~Sykes,
\textit{ibid.} \textbf{9}, 1109 (1976).

\bibitem{kesten} H.~Kesten,
{\em Percolation Theory for Mathematicians},
(Brikhauser, Boston, 1982).

\bibitem{bunde} D.~Stauffer, 
{\em Percolation} in {\em Fractals and Disordered Systems},
Eds. A.~Bunde and S.~Hauling,
(Springer, Berlin, 1991).

\bibitem{intro} D.~Stauffer, A.~Aharony, 
{\em Introduction to Percolation Theory},
(Taylor and Francis, London, 1994).

\bibitem{sahimi} M.~Sahimi,
{\em Applications of Percolation Theory},
(Taylor and Francis, London, 1994).

\bibitem{ds-rev} D.~Stauffer,
Physica A \textbf{242}, 1 (1997).

\bibitem{suding} 
P.~N.~Suding, R.~M.~Ziff,
Phys. Rev. E \textbf{60}, 275 (1999).

\bibitem{rosowsky} 
A.~Rosowsky,
Eur. Phys. J. B \textbf{15}, 77 (2000).

\bibitem{network}
M.~E.~J.~Newman, I.~Jensen, R.~M.~Ziff, 
Phys. Rev. E \textbf{65}, 021904 (2002); 
D.~S.~Callaway, M.~E.~J.~Newman, S.~H.~Strogatz, D.~J.~Watts, 
Phys. Rev. Lett. \textbf{85}, 5468 (2000); 
C.~Moore, M.~E.~J.~Newman,
Phys. Rev. E \textbf{62}, 5678 (2000); 
C.~Moore, M.~E.~J.~Newman, 
\textit{ibid.} \textbf{62}, 7059 (2000). 

\bibitem{pcd13}
S.~Galam, A.~Mauger,
Physica A \textbf{205}, 502 (1994);
P.~Grassberger,
Phys. Rev. E \textbf{67}, 036101 (2003).

\bibitem{bulg} I.~V.~Petrov, I.~I.~Stoynev, F.~V.~Babalievskii, 
J. Phys. A \textbf{24}, 4421 (1991).

\bibitem{pcdz}
S.~Galam, A.~Mauger,
Phys. Rev. E \textbf{53}, 2177 (1996);
S.~Galam, A.~Mauger,
Phys. Rev. E \textbf{56}, 322 (1997);
S.~Galam, A.~Mauger,
Eur. Phys. J. B \textbf{1}, 255 (1998).

\bibitem{marck}
S.~C.~van der Marck,
Phys. Rev. E \textbf{55}, 1228 (1997);
S.~C.~van der Marck,
\textit{ibid.} \textbf{55}, 1514 (1997);
S.~C.~van der Marck,
\textit{ibid.} \textbf{55}, 3732(E) (1997);
S.~C.~van der Marck,
Int. J. Mod. Phys. C \textbf{9}, 529 (1998);
F.~Babalivski,
Phys. Rev. E \textbf{59}, 1278 (1999).

\bibitem{newman} M.~E.~J.~Newman, R.~M.~Ziff,
Phys. Rev. E \textbf{64}, 016706 (2001);
Phys. Rev. Lett. \textbf{85}, 4104 (2000).

\bibitem{ziff} R.~M.~Ziff, 
J. Stat. Phys. \textbf{28}, 838 (1982).

\bibitem{leath} P.~L.~Leath, 
Phys. Rev. B \textbf{14}, 5046 (1976).

\bibitem{hka} J.~Hoshen, R.~Kopelman,
Phys. Rev. B \textbf{14}, 3428 (1976).

\bibitem{km-amv} K.~Malarz, A.~M.~Vidales, 
Int. J. Mod. Phys. C \textbf{9}, 147 (1998).

\bibitem{ds-aa} D.~Stauffer, A.~Aharony, 
Int. J. Mod. Phys. C \textbf{10}, 935 (1999).

\bibitem{km-sk-kk} K.~Malarz, S.~Kaczanowska, K.~Ku{\l}akowski, 
Int. J. Mod. Phys. C \textbf{13}, 1017 (2002).

\bibitem{pcNN} R.~M.~Ziff, 
Phys. Rev. Lett. \textbf{69}, 2670 (1992).

\bibitem{dd}
C.~Domb, N.~W.~Dalton,
Proc. Phys. Soc. \textbf{89}, 859 (1966).

\bibitem{wier} J.~C.~Wierman,
Phys. Rev. E \textbf{66}, 27105 (2002). 

\bibitem{solomon} K.~Malarz,
Int. J. Mod. Phys. C \textbf{13}, 1017 (2002).

\bibitem{bragg} W.~L.~Bragg, E.~J.~Williams, 
Proc. Roy. Soc. A \textbf{145}, 699 (1934).

\bibitem{huang} K.~Huang,
{\em Statistical Mechanics},
(John Willey and Sons, Inc., New York, 1963).

\bibitem{Tc_alld} S.~Galam, A.~Mauger,
Physica A \textbf{235}, 573 (1997).

\bibitem{fires}
G.~Caldarelli, \textit{at al.}, 
{\tt cond-mat/0108011};
J.~A.~M.~S.~Duarte,
Ann. Rev. Comput. Phys. \textbf{V}, 1 (1997).

\bibitem{immun} 
A.~Jimenez-Dalmaroni, H.~Hinrichsen,
{\tt cond-\-mat/\-0304113};
S.~M.~Dammer, H.~Hinrichsen,
{\tt cond-mat/0303467};
L.~M.~Sander, C.~P.~Warren, I.~M.~Sokolov,
{\tt cond-\-mat/\-0301394}.

\bibitem{porous} 
Y.~Lee \textit{at al.}, 
Phys. Rev. E \textbf{60}, 3425 (1999);
A.~M.~Vidales \textit{at al.}, 
Europhys. Lett. \textbf{36}, 259 (1996).

\bibitem{econo}
A.~Barra\~n\'on,
{\tt nlin.AO/0404009};
M.~Hohnisch, S.~Pittnauer, D.~Stauffer,
{\tt cond-\-mat/\-0308358};
D.~Makowiec, P.~Gnaci\'nski, W.~Miklaszeski,
{\tt cond-\-mat/\-0307290};
J.~Goldenberg, B.~Libai, S.~Solomon, N.~Jan, D.~Stauffer,
{\tt cond-mat/0005426}.

\bibitem{socio}
T.~Erez, S.~Moldovan, S.~Solomon,
{\tt cond-mat/0406695};
S.~Galam, A.~Mauger,
Physica A \textbf{323}, 695 (2003);
S.~Galam,
Physica A \textbf{336}, 49  (2004);
S.~Galam,
Eur. Phys. J. B \textbf{26}, 269 (2002);
D.~Stauffer,
{\tt cond-mat/0204099};
S.~Galam,
{\tt cond-mat/0204052};
A.~Proykova, D.~Stauffer,
{\tt cond-mat/0203375};
S.~Solomon, G.~Weisbuch, L.~de~Arcangelis, N.~Jan, D.~Stauffer,
Physica A \textbf{277}, 239 (2000).

\end{thebibliography}
\end{document}